\documentclass[12pt]{article}
\usepackage{graphicx}
\usepackage{epstopdf}
\usepackage{amsmath,amssymb,amsfonts,graphicx}

\begin{document}
\thispagestyle{empty}
\renewcommand{\refname}{References}

\title{\bf On the chiral separation effect in a slab} 

\author{Yurii A. Sitenko$^{1,2}$}

\date{}

\maketitle
\begin{center}
$^{1}$ Bogolyubov Institute for Theoretical Physics,\\
National Academy of Sciences of Ukraine,\\
14-b Metrologichna Street, 03680 Kyiv, Ukraine\\
$^{2}$ Institute for Theoretical Physics, University of Bern,\\
Sidlerstrasse 5, CH-3012 Bern, Switzerland

\end{center}

\begin{abstract}
We study the influence of boundaries on chiral effects in hot dense relativistic spinor matter 
in a strong magnetic field which is transverse to bounding planes. The most general set of boundary 
conditions ensuring the confinement of matter within the bounding planes is considered. We find that, 
in thermal equilibrium, the nondissipative axial current along the magnetic field is induced, depending 
on chemical potential and temperature, as well as on a choice of boundary conditions. As temperature increases 
from zero to large values, a stepwise behaviour of the axial current density as a function 
of chemical potential is changed to a smooth one; the choice of a boundary condition can facilitate either 
amplification or diminution of the chiral separation effect. This points at a significant role of boundaries 
for physical systems with hot dense magnetized relativistic spinor matter, e.g., compact stars, heavy-ion 
collisions, novel materials known as Dirac and Weyl semimetals.\end{abstract}


\bigskip

\begin{center}
\end{center}

\bigskip

\section{Introduction}

Properties of hot dense relativistic matter in a strong magnetic field are intensively studied during the last decade. An interest 
to this subject is continuously augmented in view of possible applications to diverse areas of contemporary physics, ranging from 
cosmology, astroparticle and high-energy physics to condensed-matter physics. Namely, these studies are relevant for physical 
processes in the early universe \cite{Tash}, compact astrophysical objects (neutron stars and magnetars) \cite{Char}, relativistic 
heavy-ion collisions \cite{Khar}, novel materials known as Dirac and Weyl semimetals \cite{Liu,Vaf}. Assuming that temperature 
and chemical potential, as well as the inverse magnetic length, exceed considerably the mass of a relativistic spinor matter field, 
a variety of chiral effects emerges in hot dense magnetized relativistic spinor matter, see review in \cite{Mir} and references 
therein.

So far chiral effects were mostly considered in unbounded matter, which perhaps may be relevant for cosmological applications. 
For all other applications (to astroparticle, high-energy and condensed-matter physics), an account has to be taken for the 
finiteness of physical systems, and the role of boundaries in chiral effects in bounded matter should be clearly exposed. A first 
step in this direction was undertaken in recent work \cite{Gor}, and in the present letter the results of this work will be 
extended to the case of nonzero temperature and more general boundary conditions.

\section{Choice of boundary condition}

Although the concept of quasiparticle excitations confined inside bounded-material samples is quite familiar in the context of 
condensed-matter physics, a quest for boundary conditions ensuring the confinement was initialed in particle physics in the 
context of a model description of hadrons as bags containing quarks \cite{Bog, Cho1}. Motivations for a concrete form of the 
boundary condition may differ in detail, but the key point is that the boundary condition has to guarantee the vanishing of the 
vector current of quark matter across the boundary, see \cite{Joh}. However, from this point of view, the bag boundary conditions 
proposed in \cite{Bog, Cho1} are not the most general ones. It has been rather recently realized that the most general boundary 
condition ensuring the confinenent of relativistic spinor matter within a simply-connected boundary involves four arbitrary 
parameters \cite{Bee, Wie}. An explicit form of such a condition can be given as (see \cite{Si1, Si2, Si3})
\begin{equation}
\left\{\gamma^0-\left[{\rm e}^{{\rm i}\varphi\gamma^5}\cos\theta + (\gamma^1\cos\phi + 
\gamma^2\sin\phi)\sin\theta\right]{\rm e}^{{\rm i}\tilde{\varphi}\gamma^0(\boldsymbol{\gamma}\cdot\boldsymbol{n})}\right\}
\chi(\mathbf{r})\left.
\right|_{\mathbf{r}\in \partial\Omega}=0,\label{eq1}
\end{equation}
where $\gamma^5=-{\rm i}\gamma^0\gamma^1\gamma^2\gamma^3$ ($\gamma^0$, $\gamma^1$, $\gamma^2$ and $\gamma^3$ are the generating elements 
of the Dirac-Clifford algebra), $\boldsymbol{n}$ is the unit normal to surface $\partial\Omega$ bounding spatial region $\Omega$ 
and $\chi(\mathbf{r})$ is the confined spinor matter field, $\mathbf{r} \in \Omega$; matrices $\gamma^1$ and $\gamma^2$ in (1) are 
chosen to obey condition
\begin{equation}
[\gamma^1,\,\boldsymbol{\gamma}\cdot\boldsymbol{n}]_+=
[\gamma^2,\,\boldsymbol{\gamma}\cdot\boldsymbol{n}]_+=[\gamma^1,\,\gamma^2]_+=0,\label{eq2}
\end{equation}
and the boundary parameters in (1) are chosen to vary as
\begin{equation}
-\frac{\pi}{2}<\varphi\leq\frac{\pi}{2}, \quad -\frac{\pi}{2}\leq\tilde{\varphi}<\frac{\pi}{2}, \quad 
0\leq\theta<\pi, \quad 0\leq\phi<2\pi. \label{eq3}
\end{equation}
The MIT bag boundary condition \cite{Joh},
\begin{equation}
(I+{\rm i}\boldsymbol{\gamma}\cdot\boldsymbol{n})\chi(\mathbf{r})\left.\right|_{\mathbf{r}\in \partial\Omega}=0, \label{eq4}
\end{equation}
is obtained from (1) at $\varphi=\theta=0$, $\tilde{\varphi}=-{\pi}/{2}$.

The boundary parameters in (1) can be interpreted as the self-adjoint extension parameters. The self-adjointness of the 
one-particle energy (Dirac Hamiltonian in the case of relativistic spinor matter) operator in first-quantized theory is required 
by general principles of comprehensibility and mathematical consistency, see \cite{Neu}. To put it simply, a multiple action is 
well defined for a self-adjoint operator only, allowing for the construction of functions of the operator, such as resolvent, 
evolution, heat kernel and zeta-function operators, with further implications upon second quantization.

In the case of a disconnected boundary consisting of two simply-connected components, 
$\partial\Omega=\partial\Omega^{(+)}\bigcup\partial\Omega^{(-)}$, there are in general 8 boundary parameters: 
$\varphi_{+}$, $\tilde{\varphi}_{+}$, $\theta_+$ and $\phi_+$ corresponding to $\partial\Omega^{(+)}$ and 
$\varphi_{-}$, $\tilde{\varphi}_{-}$, $\theta_{-}$ and $\phi_-$ corresponding to $\partial\Omega^{(-)}$. However, if some symmetry 
is present, then the number of boundary parameters is diminished. As in \cite{Gor}, let us consider spatial region $\Omega$ in the 
form of a slab bounded by parallel planes, $\partial\Omega^{(+)}$ and $\partial\Omega^{(-)}$, separated by distance $a$. Then the 
cases differing by the values of $\phi_+$ and $\phi_-$ are physically indistinguishable, since they are related by a rotation 
around a normal to the boundary. To avoid this unphysical degeneracy, one has to fix $\theta_+=\theta_-=0$, and the boundary 
condition takes form
\begin{equation}
\left\{\gamma^0-\exp\left[{\rm i}\left(\varphi_\pm\gamma^5\pm\tilde{\varphi}_\pm\gamma^0\gamma^z\right)\right]\right\}
\chi(\mathbf{r})\left.\right|_{z=\pm a/2}=0,\label{eq5}
\end{equation}
where coordinates $\mathbf{r}=(x,\,y,\,z)$ are chosen in such a way that $x$ and $y$ are tangential to the boundary, while $z$ is 
normal to it, and the position of $\partial\Omega^{(\pm)}$ is identified with $z=\pm a/2$. Condition (5) determines the spectrum of 
the wave number vector in the $z$-direction, $k_l$. The requirement that this spectrum be real and unambiquous yields constraint
\begin{equation}
\varphi_+=\varphi_-=\varphi, \quad \tilde{\varphi}_+=\tilde{\varphi}_-=\tilde{\varphi};\label{eq6}
\end{equation}
then the $k_l$-spectrum is determined implicitly from relation \cite{Si2, Si3}
\begin{equation}
k_l\sin\tilde{\varphi}\cos(k_l a)+(E_{k_{\bot} k_l}\cos\tilde{\varphi}-M\cos\varphi)\sin(k_l a)=0, \label{eq7}
\end{equation}
where $M$ is the mass of the spinor matter field, $E_{k_\bot k_l}$ is the energy of a one-particle state, $k_\bot$ is the absolute 
value of the wave number vector which is tangential to the boundary. In the case of the massless spinor matter field, $M=0$, 
relation (7) takes form
\begin{equation}
k_l\sin\tilde{\varphi}\cos(k_l a)+E_{k_\bot k_l}\cos\tilde{\varphi}\sin(k_l a)=0,\label{eq8}
\end{equation}
depending on $\tilde{\varphi}$ only, whereas the boundary condition  
depends on $\tilde{\varphi}$ and $\varphi$ as well:
\begin{equation}
\left\{\gamma^0-\exp\left[{\rm i}\left(\varphi\gamma^5 \pm  \tilde{\varphi}\gamma^0\gamma^z\right)\right]\right\}\chi(\mathbf{r})|_{z=\pm a/2}=0. \label{eq9}
\end{equation}

\section{Current of hot dense magnetized matter} 

Let us consider hot dense  relativistic spinor matter in the background of a static uniform magnetic field which, as in \cite{Gor}, 
is orthogonal to the slab: $\mathbf{B}=(0,\,0,\,B)$; thus the above-mentioned rotational symmetry is maintained. The use of the slab 
geometry is motivated by our aim to understand the effect of a boundary which is transverse to a magnetic field; note also that the 
slab geometry is conventional in a setup for the Casimir effect \cite{Cas1}, see review in \cite{Bor}. Assuming that field strength 
$B$, temperature $T$ and chemical potential $\mu$ are large,
\begin{equation}
|eB|>>M^2, \quad T>>M, \quad \mu>>M \label{eq10}
\end{equation}
($e$ is the charge of the matter field, natural units $\hbar=c=k_{\rm B}=1$ are used), we shall employ an approximation neglecting 
the mass of the matter field and put $M=0$ in the following. The one-particle energy spectrum is
\begin{equation}
E_{k_\bot k_l} = \pm\sqrt{k_\bot^2+k_l^2}, \quad k_\bot=\sqrt{2n|eB|}, \label{eq11}
\end{equation}
where values $n=0,\,1,\,2,\,\ldots$ correspond to the Landau levels and values of $k_l$ are determined from (8). The component of 
the axial current density which is directed along the magnetic field in the slab is defined as
\begin{eqnarray}
J^{z5}=\sum\limits_{l}\sum\limits_{n=0}^{\infty}{\rm sgn}(E_{k_\bot k_l})
\bar{\psi}_{k_\bot k_l}(\mathbf{r})\gamma^z\gamma^5\psi_{k_\bot k_l}(\mathbf{r})\nonumber \\ 
\times\Biggl(\exp\left\{\left[\,|E_{k_\bot k_l}|-{\rm sgn}(E_{k_\bot k_l})\mu\right]/T\right\}+1\Biggr)^{-1},\label{eq12}
\end{eqnarray}
where functions $\psi_{k_\bot k_l}(\mathbf{r})$ form a complete set of solutions to the stationary Dirac equation,
\begin{equation}
-{\rm i}\gamma^0\boldsymbol{\gamma}\cdot(\boldsymbol{\partial}-{\rm i}e\mathbf{A})\psi_{k_\bot k_l}(\mathbf{r})=
E_{k_\bot k_l}\psi_{k_\bot k_l}(\mathbf{r}),\label{eq13}
\end{equation}
$\bar{\psi}=\psi^{\dagger}\gamma^0$, $\mathbf{A}=\frac 12{\mathbf{B}}\times\mathbf{r}$, ${\rm sgn} (u)=1$ at $u>0$ and 
${\rm sgn}(u)=-1$ at $u<0$. Only the lowest Landau level ($n=0$) contributes to (12), thus the spectrum of the wave number vector 
in the direction of the magnetic field is
\begin{equation}
k_l=[l\pi-{\rm sgn}(E_{0k_l})\tilde{\varphi}]/a, \quad l\in\mathbb{Z}, \quad k_l>0,\label{eq14}
\end{equation}
$\mathbb{Z}$ is the set of integer numbers. Then the calculation of the sum over $l$ yields 
(details will be published elsewhere)
\begin{equation}
J^{z5}=-\frac{eB}{2\pi a}{\rm sgn}(\mu)F\left(|\mu|a + {\rm sgn}(\mu)\left[\tilde{\varphi}-
{\rm sgn}(\tilde{\varphi})\frac{\pi}{2}\right]; Ta\right),\label{eq15}
\end{equation}
where
\begin{eqnarray}
F(s;t)=\frac{s}{\pi}-\frac{1}{\pi}\int\limits_{0}^{\infty}{dv\frac{\sin(2s){\rm sinh}(\pi/t)}
{[\cos(2s)+{\rm cosh}(2v)][{\rm cosh}(\pi/t)+\cos(v/t)]}}\nonumber \\ 
+\frac{{\rm sinh} \left\{[{\rm arctan} ({\rm tan} s)]/t\right\}}{{\rm cosh}[\pi/(2t)]+{\rm cosh}
\left\{[{\rm arctan} ({\rm tan} s)]/t\right\}}.\label{eq16}
\end{eqnarray}

In view of relation
\begin{equation}
\lim\limits_{a\rightarrow \infty}\frac{1}{a}F(|\mu|a;Ta)=|\mu|/{\pi},\label{eq17}
\end{equation}
the case of a magnetic field filling the whole (infinite) space \cite{Vil,Met} is obtained from (15) as a limiting case:
\begin{equation}
\lim\limits_{a\rightarrow \infty}J^{z5}=-\frac{eB}{2\pi^2}\mu.\label{eq18}
\end{equation}
Unlike this unrealistic case, the realistic case of a magnetic field confined to a slab of finite width is temperature dependent, 
see (15) and (16). In particular, we get
\begin{equation}
\lim\limits_{T\rightarrow 0}J^{z5}=-\frac{eB}{2\pi a}{\rm sgn}(\mu)[\![[|\mu|a+{\rm sgn}(\mu)\tilde{\varphi}]/\pi+
\Theta(-\mu\tilde{\varphi})]\!] \label{eq19}
\end{equation}
and
\begin{equation}
\lim\limits_{T\rightarrow \infty}J^{z5}=-\frac{eB}{2\pi^2}\left\{\mu + 
[\tilde{\varphi}-{\rm sgn}(\tilde{\varphi})\pi/2]/a\right\}; \label{eq20}
\end{equation} 
here $[\![u]\!]$ denotes the integer part of quantity $u$ (i.e. the integer which is less than or equal to $u$) and 
$\Theta(u)=\frac{1}{2}[1+{\rm sgn}(u)]$. As follows from (15), the boundary condition that is parametrized by $\tilde{\varphi}$ 
can serve as a source which is additional to the spinor matter density: the contribution of the boundary to the axial 
current effectively enhances (at $-\frac{\pi}{2}<\tilde{\varphi}<0$) or diminishes (at $\frac{\pi}{2}>\tilde{\varphi}>0$) the 
contribution of chemical potential. Due to the boundary condition, the chiral separation effect can be nonvanishing even at 
zero chemical potential:
\begin{equation}
J^{z5}|_{\mu=0}=-\frac{eB}{2\pi a}F(\tilde{\varphi}-{\rm sgn}(\tilde{\varphi})\pi/2;Ta);\label{eq21}
\end{equation}
the latter vanishes in the limit of zero temperature, 
\begin{equation}
\lim\limits_{T\rightarrow 0}J^{z5}|_{\mu=0}=0.\label{eq22}
\end{equation}
The trivial boundary condition, $\tilde{\varphi}=-\pi/2$, gives spectrum $k_l=(l+\frac 12)\frac{\pi}{a}$ $\quad$ ($l=0,\,1,\,2,\,\ldots$), and the axial current density at zero temperature for this case was obtained in \cite{Gor},
\begin{equation}
J^{z5}\left.\right|_{T=0, \,\, \tilde{\varphi}=-\pi/2} = - \frac{eB}{2\pi a}{\rm sgn}(\mu)[\![ |\mu|a/{\pi}+1/2 ]\!].\label{eq23}
\end{equation}
The high-temperature limit in this case coincides with the limit of $a\rightarrow \infty$,
\begin{equation}
J^{z5}\left.\right|_{T=\infty, \,\, \tilde{\varphi}=-\pi/2} =-\frac{eB}{2\pi^2}\mu.\label{eq24}
\end{equation}
It should be noted that the "bosonic-type" spectrum, $k_l=l\frac{\pi}{a}$ $\quad$ ($l=0,\,1,\,2,\,\ldots$), is given by 
$\tilde{\varphi}=0$, but the axial current density is discontinuous at this point:
\begin{equation}
\lim\limits_{\tilde{\varphi}\rightarrow 0_+}J^{z5} - \lim\limits_{\tilde{\varphi}\rightarrow 0_-}J^{z5} = \frac{eB}{2\pi a}.\label{eq25}
\end{equation}

Concerning now the axial charge density,
\begin{eqnarray}
J^{05}=\sum\limits_{l}\sum\limits_{n=0}^{\infty}{\rm sgn}(E_{k_\bot k_l})
\bar{\psi}_{k_\bot k_l}(\mathbf{r})\gamma^0\gamma^5\psi_{k_\bot k_l}(\mathbf{r})\nonumber \\ 
\times\Biggl(\exp\left\{\left[|E_{k_\bot k_l}|-{\rm sgn}(E_{k_\bot k_l})\mu\right]/T\right\}+1\Biggr)^{-1},\label{eq26}
\end{eqnarray}
and the vector current density,
\begin{eqnarray}
J^z=\sum\limits_{l}\sum\limits_{n=0}^{\infty}{\rm sgn}(E_{k_\bot k_l})
\bar{\psi}_{k_\bot k_l}(\mathbf{r})\gamma^z\psi_{k_\bot k_l}(\mathbf{r})\nonumber \\ 
\times\Biggl(\exp\left\{\left[|E_{k_\bot k_l}|-{\rm sgn}(E_{k_\bot k_l})\mu\right]/T\right\}+1\Biggr)^{-1},\label{eq27}
\end{eqnarray}
one finds that these quantities are vanishing in the slab: 
\begin{equation}
J^{05}=0, \quad J^z=0.\label{eq28}
\end{equation}
This confirms the conclusion of \cite{Gor} about the absence of the chiral magnetic effect in the slab; note also that the chiral 
magnetic effect is doubted in the recent holographic study \cite{Jim}.

\section{Conclusion}

We have considered the influence of boundaries on chiral effects in hot dense magnetized relativistic spinor matter. In the 
absence of boundaries these effects include the chiral separation effect which is characterized by the nondissipative axial 
current along the magnetic field \cite{Vil,Met} and the chiral magnetic effect which is characterized by the nondissipative vector 
current in the same direction \cite{Fuk}; both currents are temperature-independent. As boundaries are introduced and the matter 
volume is shrinked to a slab which is transverse to the magnetic field, the fate of these currents is different. The axial current 
stays on, becoming dependent on temperature and on a choice of boundary conditions, see (15) and (16); as temperature increases 
from zero to large values, a stepwise behaviour of the axial current density as a function of chemical potential is changed to a 
smooth one, see (19) and (20). The vector current is extinct, that is due to the boundary conditions confining matter inside the 
slab. Since the axial charge is not induced in the slab as well, see (28), a motivation on the grounds of a microscopic theory for 
the introduction of the chiral chemical potential is lacking, and the chiral magnetic effect is absent in the slab, as was pointed 
first in \cite{Gor}.

Let us comment in more detail on the issue of boundary conditions in the slab geometry. In the case of the quantized 
electromagnetic matter field, a choice of boundary conditions is motivated by material properties of the bounding plates, and 
the conventional Casimir effect is different for different boundary conditions. For instance, it is attractive between 
ideal-metal plates (i.e. made of material with an infinitely large magnitude of the dielectric permittivity), as well as 
between plates made of material with an infinitely large magnitude of the magnetic permeability; meanwhile, it is repulsive 
between an ideal-metal plate and an infinitely permeable one, see, e.g., \cite{Bor}. Thus, even if the material of plates is 
unknown for some reasons, it can in principle be determined by measuring the Casimir force or other physical quantities.

Namely such a situation happens in the case of the quantized spinor matter field, when nothing can be said about the ``material'' 
of boundaries, other than to admit that this ``material'' is impenetrable for spinor matter. The ``material'' properties of the 
boundaries are encoded in the values of the boundary (self-adjoint extension) parameters. However, for zero temperature and zero 
chemical potential, the condition of confinement within the ``material'' boundaries (i.e. the absence of the vector current of 
spinor matter across the boundaries) is sufficient in itself: the physical quantity, Casimir force, is independent of the boundary 
parameters \cite{Si1,Si2,Si3}. We show in the present letter that this is in general changed when, at least, either temperature 
or chemical potential becomes nonzero: the physical quantity, axial current, becomes dependent on the boundary parameter, 
$\tilde{\varphi}$. The confining boundary condition on the sides of the slab creates a whole one-parameter family of possible 
standing waves with the wave number vector, $k_l$, given by (14). Each member of the family is allowable on an equal footing as is 
a standing wave with $k_l=(l+\frac 12)\frac{\pi}{a}$ $\quad$ ($l=0,\,1,\,2,\,\ldots$). Only a standing wave with $k_l=l\frac{\pi}{a}$ $\quad$ 
($l=0,\,1,\,2,\,\ldots$) is not allowed due to an ambiguity of the axial current, see (25); this can be understood as 
a physical unacceptability of fermionic eigefunctions with bosonic eigenvalues. 

Thus, the result of \cite{Gor} about the stepwise behaviour of the axial current density at zero temperature, see (23), is not a 
definite physical prediction to be validated in experiments; the whole pattern of steps can be shifted either to the right 
or to the left by less than a half of the step width, see (19). Another distinctive feature is the possible persistence of the 
chiral separation effect at zero chemical potential and nonzero temperature, see (21). Perhaps, it is for the first time that such 
a mathematical entity as the self-adjoint extension parameter, $\tilde{\varphi}$, is to be determined experimentally (maybe, at 
least as an event-by-event fluctuation). It would be interesting to verify this in table-top experiments with slabs of Dirac or 
Weyl semimetals in a magnetic field.

\section*{Acknowledgments}

The work was supported by the National Academy of Sciences
of Ukraine (project No.0112U000054), by the
Program of Fundamental Research of the Department of Physics and
Astronomy of the National Academy of Sciences of Ukraine (project
No.0112U000056) and by the ICTP -- SEENET-MTP grant PRJ-09
``Strings and Cosmology''.


\begin{thebibliography}{99}

\bibitem{Tash}%
H. Tashiro, T. Vachaspati and A. Vilenkin, Phys. Rev. D \textbf{86},
105033 (2012).

\bibitem{Char}%
J. Charbonneau and A. Zhitnitsky, J. Cosmol. Astropart. Phys. \textbf{1008}, 010 (2010).

\bibitem{Khar}%
D. E. Kharzeev, Progr. Part. Nucl. Phys. {\bf 75}, 133 (2014).

\bibitem{Liu}%
Z. K. Liu, B. Zhou, Y. Zhang, Z. J. Wang, H. M. Weng, D. Prabhakaran, S. -K. Mo, Z. X. Chen, Z. Fang, X. Dai, Z. Hussain and Y. L. Chen, Science {\bf 343}, 864 (2014).

\bibitem{Vaf}%
O. Vafek and A. Vishwanath, Ann. Rev. Cond. Mat. Phys. {\bf 5}, 83 (2014).

\bibitem{Mir}%
V. A. Miransky and I. A. Shovkovy, Phys. Rept. \textbf{576}, 1 (2015).

\bibitem{Gor}%
E. V. Gorbar, V. A. Miransky, I. A. Shovkovy and P. O. Sukhachov, Phys. Rev. B {\bf 92}, 245440 (2015).

\bibitem{Bog}%
P. N. Bogolioubov, Ann. Inst. Henri Poincare A \textbf{8}, 163 (1968).

\bibitem{Cho1}
A. Chodos, R. L. Jaffe, K. Johnson, C. B. Thorn and V. Weisskopf, Phys. Rev. D {\bf 9}, 3471 (1974).

\bibitem{Joh}
K. Johnson, Acta  Phys. Polon. B {\bf 6}, 865 (1975).

\bibitem{Bee}
A. R. Akhmerov and C. W. J. Beenakker, Phys. Rev. B {\bf 77},
085423 (2008).

\bibitem{Wie}
M. H. Al-Hashimi and U. -J. Wiese, Annals Phys. {\bf 327}, 1
(2012).

\bibitem{Si1}
Yu. A. Sitenko, Phys. Rev. D {\bf 91}, 085012 (2015).

\bibitem{Si2}%
Yu. A. Sitenko and S. A. Yushchenko, Intern. J. Mod. Phys. A \textbf{30}, 1550184 (2015).

\bibitem{Si3}%
Yu. A. Sitenko, J. Phys.: Conf. Series \textbf{670}, 012048 (2016).

\bibitem{Neu}
J. von Neumann, {\it Mathematische Grundlagen der Quantummechanik}
(Springer, Berlin, 1932).

\bibitem{Cas1}  H. B. G. Casimir, Proc. Kon. Ned. Akad. Wetenschap
B {\bf 51}, 793 (1948).

\bibitem{Bor} 
M. Bordag, G. L. Klimchitskaya, U. Mohideen and V. M. Mostepanenko,
{\it Advances in the Casimir Effect} (Oxford University Press, Oxford, 2009).

\bibitem{Vil}
A. Vilenkin, Phys. Rev. D \textbf{22}, 3080 (1980).

\bibitem{Met}%
M. A. Metlitski and A. R. Zhitnitsky, Phys. Rev. D \textbf{72}, 045011 (2005).

\bibitem{Jim}
A. Jimenez-Alba, K. Landsteiner and L. Melgar, Phys. Rev. D \textbf{90}, 126004 (2014).

\bibitem{Fuk}
K. Fukushima, D. E. Kharzeev and H. J. Warringa, Phys. Rev. D \textbf{78}, 074033 (2008).


\end{thebibliography}
\end{document}